\documentclass[preprint,12pt,3p]{elsarticle}

\usepackage{graphics}
\usepackage{epstopdf}
\usepackage{amsmath}
\usepackage{amssymb}

\journal{Annals of Physics}

\begin{document}

\begin{frontmatter}

\title{Kerr-Newman-NUT-Kiselev black holes in Rastall theory of gravity and Kerr/CFT Correspondence}

\author[label1]{Muhammad F. A. R. Sakti\corref{cor1}}
\address[label1]{Theoretical Physics Laboratory, THEPI Division, Institut Teknologi Bandung, Jl. Ganesha 10 Bandung, 40132, Indonesia}
\address[label2]{Indonesia Center for Theoretical and Mathematical Physics (ICTMP), Institut Teknologi Bandung, Jl. Ganesha 10 Bandung, 40132, Indonesia\fnref{label4}}

\cortext[cor1]{Corresponding author}

\ead{m.fitrah@students.itb.ac.id}

\author[label1,label2]{Agus Suroso}
\ead{agussuroso@fi.itb.ac.id}

\author[label1,label2]{Freddy P. Zen}
\ead{fpzen@fi.itb.ac.id}

\begin{abstract}
We present a new twisted rotating black hole solution by performing Demia{\'n}ski-Newman-Janis algorithm to the electrically and dyonically charged black hole with quintessence in Rastall theory of gravity. Using our black hole solution, we argue that Rastall gravity is not equivalent with Einstein gravity. For further explanation, the black hole properties such as the horizon and ergosphere are studied for which there are some different properties for those theories. Some thermodynamic properties of the black hole solution are also discussed. At the end, considering the Kerr/CFT correspondence is valid for our black hole solution, the central charge from the CFT of this extremal solution is derived.
\end{abstract}

\begin{keyword}
Demia{\'n}ski-Newman-Janis algorithm, Rastall gravity, thermodynamics, Kerr/CFT correspondence
\end{keyword}

\end{frontmatter}


\section{Introduction}
\label{sec:intro}
The Einstein theory of gravity till nowadays has been confirmed by some observational evidence in our universe. This theory considers the covariant conservation of the energy-momentum tensor. However, the Einstein theory is also believed as a certain case of some extended theories of gravity. Since the first time it was formulated, people have been working for some other alternative theories and developing several modified theories of gravity. One of the promising general theories of relativity was introduced by Rastall \cite{Rastall1972,Rastall1976}. Within this theory the usual conservation law of the energy momentum tensor is not obeyed or on the other hand, the energy-momentum tensor is not conservative ($ \nabla_\mu T^{\mu \nu} \neq 0 $). The usual conservation law, expressed by the null covariant derivative of the energy-momentum tensor, is questioned whether it is valid generally in curved space-time or only in certain cases since this theory is tested only in a weak gravitational field limit or in the flat Minkowski space–time. It is considered a non-minimal coupling of the matter field to space-time geometry with the coupling constant $ \kappa\lambda $ or Rastall parameter. This quantity quantifies the deviation from the Einstein theory of gravity. When Rastall parameter vanishes, it produces the Einstein theory of gravity. Furthermore, in \cite{Fisher2019,Moraes2019} it is argued that Rastall gravity is a special case of $ f(R,T) $ gravity.

Rastall theory of gravity might be interpreted as a direct accomplishment of the Mach principle which suggests that the inertial properties of a mass distribution are determined by the distribution of mass-energy in the external space-time \cite{Vladimir2006}. Thus, the source of gravitation, either mass-density or the elements of energy-momentum tensor, depends on the gravitational tensor. We can find also some recent research work that use this gravitational theory to explain the cosmological aspects such as the accelerating expansion of the universe and the inflationary problems \cite{Campos2013,Fabris2012,Moradpour2016,Rawaf1996,Batista2012,Carames2014,Salako2016,Smalley1974,Smalley1975,Wolf1986,Moradpour2017}. In the smaller scale, some astrophysical configurations are investigated by performing this theory. For instance, the investigation of perfect fluid spheres, compact stars, neutron stars, black holes, and wormholes is done in the context of Rastall gravity \cite{Hansraj2019,Abbas2018,Abbas2019,Oliveira2015,Moradpour2017a,Moradpour2016a,Oliveira2016,Bronnikov2016,Hadyan2019}. Corresponding to black hole solutions in Rastall gravity, a fascinating non-commutative inspired black hole solution is obtained in \cite{Ma2017}. In addition, the black hole solution with the source of a Gaussian matter distribution is obtained in \cite{Spalluci2018}.  Several extensions of black hole solution in Rastall gravity are further investigated in \cite{Heydarzade2017,Heydarzade2017Cad,Kumar2018,Xu2018,KumarShadow2018}. Besides that, the thermodynamic properties of black hole solutions in the Rastall gravity are discussed in \cite{Lobo2018}.

Recently, it is claimed by Visser \cite{Visser2018} that Rastall theory of gravity is equivalent with Einstein theory. However, Darabi \textit{et al.} \cite{Corda2018} compare these two gravitational theories and summarize that Visser's conclusion is not correct. The argument in \cite{Corda2018}, indeed, supports Rastall theory of gravity for which this theory is still an open theory comparing to the usual general relativity. Henceforth, this theory may face the challenges of cosmological observations as the general relativity.

In recent years, it is found that our universe experiences accelerated expansion due to the existence of dark energy that fills more than $ 70\% $ matters of our universe. A promising model of dark energy is the quintessence. Quintessence model is a model of a scalar field that governs the pressure of our universe to be negative then makes the accelerated expansion \cite{Copeland2006}. The study involving the scalar field in some frame of modified gravity theories in a cosmological scale can be discovered in \cite{Arianto2007,Arianto2008,Zen2009,Arianto2010,Feranie2010,Arianto2011,Suroso2013,Suroso2015,Getbogi2016}.  The existence of quintessence possibly causes the change of the structure of some locally astrophysical manifestations of it such as black holes. First investigation of the quintessential effect on the black hole originally is studied by Kiselev \cite{Kiselev2003}. It is found that the black hole solution is dependent on the quintessential intensity $ \alpha $ and equation of state $ \omega $ of this type of field. The equation of state of the quintessence could be varied that depends on the domination of the matter. The value range of the equation of state $ \omega $ is $ -1/3 < \omega <0 $ which makes the black hole solutions is asymptotically flat and $ -1 < \omega < -1/3 $ which represents the accelerating expansion of universe. But it is still possible to set $ \omega $ for some special cases. In addition, black hole solution with the quintessential matter in Lovelock gravity is investigated in \cite{Maharaj2017} while in Gauss-Bonnet gravity can be found in \cite{Lee2018}. 

In the following, we derive an electrically and dyonically charged black hole solution in Rastall theory of gravity with the existence quintessential matter. We add a magnetic charge in this solution rather than adding the electric charge only \cite{Heydarzade2017} to make it more general and provide an interesting feature related to the magnetic monopole. In addition, the twisted rotating solution is more fascinating because there exist spin $ a $ and the NUT charge. Regarding NUT charge as a twisting parameter of the surrounding space-time, one can find the explanation in \cite{Badawi2006}. Another definition of this parameter can be discovered in \cite{Nouri1997} for which it tells that NUT charge represents a gravo-magnetic monopole parameter of the central mass. Herein we employ the Demia{\'n}ski-Newman-Janis algorithm \cite{Erbin2015,Erbin2016}, the more general one than the Newman-Janis algorithm, to insert the spin $ a $ together with the NUT charge $ n $. Hence we present a novel solution so-called the Kerr-Newman-NUT-Kiselev black hole solution in Rastall gravity. The solution that we obtain will be highly favorable to study the astrophysical aspects of the magnetic monopole in relation with the quintessential matter. We also give some arguments to say that Rastall gravity is not equivalent with Einstein gravity. Some properties and thermodynamic quantities of this solution are also investigated. It will be interesting to make similar studies in other modified gravity models for example $f(R)$ gravity \cite{Buchdahl1970}, Ho{\v r}ava-Lifshitz gravity \cite{PetrHorava2009} and bumblebee gravity \cite{Engelsetal2016,Maluvetal2015,KanziSakalli2019,JesusSantos2019,OliveiraAlmeida2019,AliSakalli2019,CapeloParamos2015,GuiomarParamos2014,CasanaSantos2018} which have been getting significant attention in many different aspects.

In a recent investigation, the microscopic entropy of black holes may be derived by governing the Kerr/CFT correspondence which is originally proposed in \cite{Guica2009}. This entropy merely agrees with the Bekenstein-Hawking entropy that obeys the area law of black holes \cite{Hartman2009}-\cite{SaktiEPJP}. The famous Cardy entropy formula is applied in this fashion where it is a function of the central charge and conformal temperature. The central charge comes from the central term in which it comes from the Dirac bracket of the canonically conserved charges associated with the non-trivial diffeomorphisms of the near-horizon region. This conserved charge corresponds with the theory that is implemented in \cite{Barnich2002,Barnich2008} for Einstein-Maxwell system and for several modified gravitational theories is reviewed in \cite{Adami2017} and the references therein. However, for Rastall theory of gravity, the asymptotic conserved charged has not been derived. Because of this, we assume that the Kerr/CFT correspondence is true in this theory to obtain the central charge of the solution that we find from the Bekenstein-Hawking entropy.

We set up the remaining parts of the paper as follows. In section 2, we derive the spherically symmetric charged black hole solution in Rastall gravity. After that, to find the twisted rotating solution, Demia{\'n}ski -Newman-Janis algorithm is implemented. Moreover, we give some comments regarding the equivalency between Rastall gravity and Einstein gravity. In section 3, we study the horizon and ergosphere of the black hole solution. In the next section, several thermodynamic properties are derived. Then the entropy from the previous section is used to find the central charge of the black hole solution. Finally, the summary is provided in the next section. The resulting solution can be used to argue that our solution in Rastall gravity is not equivalent with Einstein gravity.

\section{Kerr-Newman-NUT-Kiselev black hole in Rastall gravity}
\label{sec:Kerr-Newman-NUT}
Rotational parameter or spin $ a $ and twisting parameter or NUT charge $ n $ can be tucked into the spherically symmetric solution using Demia{\'n}ski-Newman-Janis algorithm \cite{Erbin2016}. This is the extension of Newman-Janis algorithm. We employ this algorithm to the dyonic Reissner-Nordstr{\"o}m black hole surrounded by the quintessence in Rastall theory of gravity.
\subsection{Dyonic Reissner-Nordstr{\"o}m black hole with Quintessence in Rastall gravity}
\label{subsec:dyonic}
Based on Rastall's theory \cite{Rastall1972,Rastall1976}, the energy-momentum tensor is not conserved and its covariant derivative is proportional to the derivative of Ricci scalar with the coupling constant $ \lambda $, namely Rastall parameter that can be written as
\begin{eqnarray}
\nabla_\mu T^{\mu\nu} = \lambda \nabla^\nu R. \label{eq:nonconservedmatter}\
\end{eqnarray}
For the flat Minkowski space-time or specifically in a weak gravitational field limit, this will reduce to the usual conservation law. So the Einstein field equation will be modified as
\begin{eqnarray}
G_{\mu\nu}+\kappa\lambda g_{\mu\nu} R= \kappa T_{\mu\nu}, \label{eq:rastalleinstein}
\end{eqnarray}
where $ \kappa = 8\pi G $ and $ G $ is the Newtonian gravitational coupling constant.

In order to obtain spherically symmetric black hole solutions, we consider the general space-time metric in the standard Schwarzschild coordinates as
\begin{eqnarray}
ds^2 = -f(r)dt^2+f(r)^{-1}dr^2+r^2 d\Omega^2 ,\label{eq:sphersymgen}
\end{eqnarray}
where $ f(r) $ is a function depending on radial coordinate only and two-dimensional sphere $ d\Omega^2 = d\theta ^2 + \sin^2\theta d\phi^2 $. Using metric (\ref{eq:sphersymgen}) to the r.h.s. of Eq. (\ref{eq:rastalleinstein}), the non-vanishing components of gravitational tensor are given by
\begin{eqnarray}
&&G^0_0 + \kappa \lambda R = \frac{1}{r^2}(f'r-1+f)+ \kappa \lambda R, ~~ G^1_1 = G^0_0, \\
&&G^2_2 + \kappa \lambda R = \frac{1}{r^2}\left(f'r+\frac{1}{2}f''r^2 \right)+ \kappa \lambda R, ~~ G^3_3 = G^2_2, \label{eq:rastallcomp}\
\end{eqnarray}
and the Ricci scalar reads as
\begin{eqnarray}
R= -\frac{1}{r^2}\left(r^2 f'' +4rf' - 2 +2f \right),
\end{eqnarray}
in which the prime represents the derivative with respect to the radial coordinate $ r $.

We have two kinds of matter in this case which are the electromagnetic field ($ E_{\mu\nu} $) and scalar field of quintessence ($ Q_{\mu\nu} $). As derived in \cite{Kiselev2003}, the non-zero components of the energy-momentum tensor of the quintessence are
\begin{eqnarray}
Q^0_0 = Q ^1_1 = -\rho_q(r), ~~~ Q^2_2 = Q^3_3 = \frac{1}{2}(3\omega+1)\rho_q(r). \label{eq:quinenermom} \
\end{eqnarray}
$ \omega $ is the parameter of equation of state of the quintessence to determine matter domination of the solution. Nonetheless, there is an argument from Visser \cite{VisserQuint2019} which tells that Kiselev solution does not represent quintessence with isotropic pressure. However, Kiselev derives the solution by assuming the isotropic averaging over the angles of the space-like matter tensor component. The calculations end up with an equation of state like the quintessence in cosmology.

Then the energy-momentum tensor of the electromagnetic field is given by
\begin{eqnarray}
E_{\mu \nu} = \frac{2}{\kappa}\left(F_{\mu \gamma}F^\gamma_\nu - \frac{1}{4}g_{\mu\nu}F^{\gamma \beta}F_{\gamma \beta} \right),
\end{eqnarray}
where $ F_{\mu\nu} = \partial_\mu A_\nu - \partial_\nu A_\mu $ is the trace-free electromagnetic field tensor. Using Bianchi identity and the variational principle over the electromagnetic potential, we arrive at
\begin{eqnarray}
\partial_{[\sigma}F_{\mu\nu]} =0, ~~~~~\partial_\mu (\sqrt{-\bar{g}} F^{\mu\nu}) =0. \label{eq:maxwellbianchi}
\end{eqnarray}
Regarding the spherical symmetry exists in the space-time metric (\ref{eq:sphersymgen}), it is imposed the only non-vanishing components of the electromagnetic field tensor $ F^{\mu\nu} $ to be $ F^{01}= -F^{10} $ and $ F^{23}= -F^{32} $. We impose more non-zero component of the electromagnetic field tensor because we wish to obtain a more general solution than in \cite{Heydarzade2017} which only contains the electric charge. By performing this assumption to Eq. (\ref{eq:maxwellbianchi}) then, the non-vanishing electromagnetic potential is given by
\begin{eqnarray}
A_\mu dx^\mu = -\frac{e}{r}dt +g \cos\theta d\phi, \label{eq:elecpot} \
\end{eqnarray}
where $ e, g $ are defined as electric and magnetic charges, respectively. The non-zero components of the electromagnetic energy-momentum tensor are given by
\begin{eqnarray}
E^0_0 = E^1_1 = -\frac{1}{\kappa}\left( \frac{e^2}{r^4}+\frac{g^2}{r^4}\right), ~~~~~ E^2_2 = E^3_3 = \frac{1}{\kappa}\left( \frac{e^2}{r^4}+\frac{g^2}{r^4}\right). \label{eq:maxwellenermom}
\end{eqnarray}

Making use of Eqs. (\ref{eq:rastallcomp}), (\ref{eq:quinenermom}) and (\ref{eq:maxwellenermom}) to Eq. (\ref{eq:rastalleinstein}), we may derive two equations that are given by
\begin{equation}
\frac{1}{r^2}(f'r-1+f) -\frac{\kappa \lambda}{r^2}\left(r^2 f'' +4rf' - 2 +2f \right)  = -\rho_q- \frac{e^2}{r^4}-\frac{g^2}{r^4}, \label{eq:solving1}\
\end{equation}
\begin{equation}
\frac{1}{r^2}\left(f'r+\frac{1}{2}f''r^2 \right) -\frac{\kappa \lambda}{r^2}\left(r^2 f'' +4rf' - 2 +2f \right) = \frac{1}{2}\rho_q(3\omega+1) + \frac{e^2}{r^4}+\frac{g^2}{r^4} . \label{eq:solving2}\
\end{equation}
Now Eqs. (\ref{eq:solving1}) and (\ref{eq:solving2}) are able to be solved. The solution to these equations is
\begin{eqnarray}
f(r) = 1- \frac{2M}{r}+ \frac{e^2+g^2}{r^2}- \alpha r^{-\frac{1+3\omega - 6\kappa\lambda(1+\omega)}{1-3\kappa\lambda(1+\omega)}} , \label{eq:rastallpart1}
\end{eqnarray}
and the quintessence energy density that reads as
\begin{eqnarray}
\rho(r)=\frac{-2W_s \alpha}{\kappa r^{K_s}} , \
\end{eqnarray}
where
\begin{eqnarray}
W_s = -\frac{(1-4\kappa\lambda)[\kappa\lambda(1+\omega)-\omega]}{[1-3\kappa \lambda (1+\omega)]^2}, ~~~~~ K_s = \frac{3(1+\omega)-12\kappa\lambda(1+\omega)}{1-3\kappa\lambda(1+\omega)}. \nonumber\
\end{eqnarray}
Regarding $ \rho_q(r)\geq 0 $, hence we have $ W_s \alpha \leq 0 $. On the other hand, for $ W_s >0 $ the quintessential intensity $ \alpha<0 $ and \textit{vice versa}. This is the dyonic Reissner-Nordstr{\"o}m black hole with quintessence in Rastall gravity that generalizes the solution obtained in \cite{Heydarzade2017} because of the presence of the magnetic charge $ g $. The dyonic black hole solutions provide special interest as it allows one to study spherical configuration with magnetic monopole \cite{Gibbons1977,LeeNairWeinberg1992}. 

\subsection{Rotating, twisting, charged black hole with quintessence in Rastall theory of gravity}
\label{subsec:rotating}

We will derive the more general solution than the electrically charged rotating black hole solution surrounded by the quintessence in Rastall gravity that is obtained in \cite{Xu2018}. Demia{\'n}ski-Newman-Janis algorithm, the extended Newman-Janis algorithm, will be performed to the solution that we have gained in the previous section. It is firstly proposed in \cite{Demianski1972} and exposed in detail in \cite{Erbin2016}. It is calculated in \cite{Erbin2016} that this algorithm satisfies the Einstein equation. In addition, Newman-Janis algorithm is valid to derive Kerr-Newman-AdS black hole solution surrounded by the quintessential matter in Rastall gravity \cite{Xu2018}. So it denotes that this algorithm may be applied in our spherically symmetric solution. In this manner, the tetrad formalism is not applied but we make use of only the complex coordinate transformation as shown in \cite{Erbin2016}. Starting the algorithm, the spherically symmetric metric (\ref{eq:sphersymgen}) is written in more general form as
\begin{eqnarray}
ds^2=-f_t(r)dt^2+f_r(r)dr^2+ f_s(r) d\Omega^2 , \label{eq:generalsphericallysymmetric}\
\end{eqnarray}
where here $ d\Omega^2 = d\theta ^2 + H^2(\theta) d\phi^2 $ and 
\begin{equation}
    H(\theta) = \left\{\begin{array}{lr}
        \sin\theta, k=1, \\
        1, k=0,\\
        \sinh\theta, k=-1.
        \end{array}\right.  
\end{equation}
The constant $ k $ is defined as the sign of the surface curvature. The electromagnetic vector potential is written in the form
\begin{eqnarray}
A_\mu dx^\mu = f_a dt + g\cos\theta d\phi .\
\end{eqnarray}
However because of the vector potential (\ref{eq:elecpot}) contains no effect from the quintessence and Rastall parameter, the later potential will remain the same as the usual Kerr-Newman-NUT black holes \cite{Sakti2018}. So we stress the derivation only for the metric. 

Using this null coordinate transformation
\begin{eqnarray}
dt = du + \sqrt{\frac{f_r}{f_t}}dr, \
\end{eqnarray}
makes the metric (\ref{eq:generalsphericallysymmetric}) becomes
\begin{eqnarray}
ds^2 = -f_t du^2 -2\sqrt{f_t f_r} du dr + f_s d\Omega^2 . 
\end{eqnarray}
After gaining the above form, we need to make use of the complex coordinate to the coordinate $ u $ and $ r $, that is given by
\begin{eqnarray}
r = \hat{r} + i F(\theta), ~~~~~
u = \hat{u} + i G(\theta), \label{eq:r,u}
\end{eqnarray}
where $ \hat{u},\hat{r} \in \mathbb{R}$, and $ F(\theta), G(\theta) $ are two arbitrary functions. To find the twisting and rotating solution, $ F(\theta)$ and $ G(\theta) $ have form as
\begin{eqnarray}
 F(\theta) = n - a \cos\theta, ~~~~~
 G(\theta) = a \cos\theta - 2n \textrm{ln}(\sin\theta), \label{eq:FGfunction}
\end{eqnarray}
where $ a, n $ are the spin and NUT charge, respectively. Along with the coordinate transformation, the coordinate $ r $ and mass $ M $ are required to be complexified by obeying 
\begin{eqnarray}
\hat{r} \rightarrow \frac{1}{2}(\hat{r}+\bar{r}) = \textrm{Re}(\hat{r}), ~~~~
\hat{r}^2 \rightarrow  |\hat{r}|^2, ~~~~
\frac{M}{\hat{r}} \rightarrow  \frac{1}{2}\left( \frac{M}{\hat{r}}+\frac{\bar{M}}{\bar{r}} \right) = \frac{M\bar{r}+\bar{M}\hat{r}}{|\hat{r}|^2}, \
\end{eqnarray}
where the mass is complexified to $ M \rightarrow M+in $. Note that $ \bar{r} $ is the conjugate of $ \hat{r} $ and  $ \bar{M} $ is the conjugate of $ M $. Next, the differential form of (\ref{eq:r,u}) is
\begin{eqnarray}
dr = d\hat{r} + i F'(\theta)d\theta, ~~~~~
du = d\hat{u} + i G'(\theta)d\theta , \label{eq:dr,du}\
\end{eqnarray}
where the prime on the function $ F(\theta),G(\theta) $ denotes the derivative respects to coordinate $ \theta $. Then we have to use Giampieri's ansatz \cite{Giampieri1990} on the angular coordinate, i.e.
\begin{eqnarray}
id\theta = H(\theta) d\phi .\
\end{eqnarray}
So, Eq. (\ref{eq:dr,du}) becomes
\begin{eqnarray}
dr = d\hat{r} + F'(\theta)H(\theta) d\phi, ~~~~~
du = d\hat{u} +  G'(\theta)H(\theta) d\phi . \label{eq:dr1,du1}\
\end{eqnarray}

Hence we may replace $ f_t, f_r, f_s $ by $ \bar{f}_t, \bar{f}_r, \bar{f}_s $ where it shows that the metric functions are transformed by complex coordinate now. Finally we obtain the twisting and rotating solution in Eddington-Finkelstein coordinates as follows
\begin{eqnarray}
ds^2 &=& - 2\sqrt{\bar{f}_t \bar{f}_r}(d\hat{u} d\hat{r}+G'H d\hat{r}d\phi) -2(\bar{f}_t G'H  +\sqrt{\bar{f}_t \bar{f}_r}F'H)d\hat{u}d\phi \nonumber\\
&-&\bar{f}_t d\hat{u}^2+\bar{f}_s d\theta^2 -(\bar{f}_tG'^2 H^2 + 2\sqrt{\bar{f}_t \bar{f}_r}F'G'H^2 -\bar{f}_s H^2)d\phi^2 .\label{eq:nutspinkriskalgen}
\end{eqnarray}
In order to find the solution in Boyer-Linquist coordinates, we need to transform metric (\ref{eq:nutspinkriskalgen}) with the following coordinates transformation
\begin{eqnarray}
d\hat{u} = d\hat{t} - g(\hat{r}) d\hat{r}, ~~~~~
 d\phi = d\varphi - h(\hat{r}) d\hat{r}, \label{eq:integrabletheta}
\end{eqnarray} 
where
\begin{eqnarray}
g(\hat{r}) = \frac{(\bar{f}_t \bar{f}_r)^{-\frac{1}{2}}\bar{f}_s - F'G'}{\frac{\bar{f}_s}{\bar{f}_r}+F'^2}, ~~~
h(\hat{r}) = \frac{F'}{H\left(\frac{\bar{f}_s}{\bar{f}_r}+F'^2\right)}.\label{eq:integrable}
\end{eqnarray}
It is worth mentioning that the functions $ g(\hat{r}),h(\hat{r}) $ cannot be dependent of coordinate $ \theta $ because it can make the transformations (\ref{eq:integrabletheta}) unintegrable \cite{Erbin2016}.  At the end, after performing (\ref{eq:integrabletheta}), the general twisted rotating metric in Boyer-Linquist coordinates reads as
\begin{eqnarray}
ds^2 = -\bar{f}_t (dt+\chi H d\varphi)^2 + \frac{\bar{f}_s}{\Delta}dr^2 + \bar{f}_s(d\theta^2 + \sigma^2 H^2 d\varphi^2), \label{eq:boyerlinquistnutspingeneral} \
\end{eqnarray}
where
\begin{eqnarray}
\chi = G' + \sqrt{\frac{\bar{f}_r}{\bar{f}_t}}F', \sigma^2 = 1+ \frac{\bar{f}_r}{\bar{f}_s}F'^2, \Delta = \frac{\bar{f}_s}{\bar{f}_r}+F'^2, \nonumber\
\end{eqnarray}
and the hat $ (\hat{~}) $ on $ t,r $ has been omitted. This metric is invariant under transformation $ (F,G,\varphi) \rightarrow -(F,G,\varphi)$.

Regarding the dyonic Reissner-Nordstr{\"o}m solution with the quintessence in Rastall gravity with $ f(r)$ that is given in Eq. (\ref{eq:rastallpart1}), we obtain the Kerr-Newman-NUT-Kiselev black hole solution in Rastall gravity that is given by
\begin{equation}
ds^2 = -\frac{\Delta}{\rho^2}\left[dt - (a\sin^2\theta +2n\cos\theta)d\varphi \right]^2+ \frac{\rho^2}{\Delta}dr^2+\rho ^2 d\theta ^2 +\frac{\sin^2\theta}{\rho ^2}\left[adt-(r^2+a^2+n^2)d\varphi \right]^2 , \label{eq:metricresult}
\end{equation}
where
\begin{eqnarray}
\Delta = r^2 - 2Mr+a^2+e^2+g^2-n^2 -\alpha r^\upsilon, ~~
\upsilon = \frac{1-3\omega}{1-3\kappa \lambda(1+\omega)}, ~~~
\rho^2 = r^2 + (n-a\cos\theta)^2. \nonumber\
\end{eqnarray}
The more general twisting and rotating solution can be gained by governing the different function of $ F(\theta) $ and $ G(\theta) $ but it is not implemented in \cite{Erbin2016}. By performing
\begin{eqnarray}
 F(\theta) = -n - a \cos\theta, ~~~~~ G(\theta) = a \cos\theta + 2n \textrm{ln}(\sin\theta) -2n \textrm{ln}\left( \tan \frac{\theta}{2} \right), \label{eq:FGfunction1}
\end{eqnarray}
we find $ g(\hat{r})=[r^2 +(a+n)^2]\Delta^{-1} $ and $ h(\hat{r}) = a\Delta^{-1} $, so our coordinate transformation (\ref{eq:integrable}) is integrable. Then the resulting metric reads as
\begin{eqnarray}
ds^2 &=& -\frac{\Delta}{\rho^2}\left[dt - \{ a\sin^2\theta +2n(1-\cos\theta) \} d\varphi \right]^2+ \frac{\rho^2}{\Delta}dr^2 \nonumber\\
& &+\rho ^2 d\theta ^2 +\frac{\sin^2\theta}{\rho ^2}\left[adt-\{r^2+(a+n)^2\}d\varphi \right]^2 , \label{eq:metricresult1}
\end{eqnarray}
where $\rho^2 = r^2 + (n+a\cos\theta)^2 $. The electromagnetic potential related to the metric (\ref{eq:metricresult1}) is given by \cite{Podolsky2009}
\begin{eqnarray}
A_\mu dx^\mu &=& \frac{-er\left[a dt- \left\{ a^2\sin^2\theta +2an(1-\cos\theta) \right\} d\varphi \right]}{a\rho ^2}  \nonumber\\
&-& \frac{g(n+a\cos\theta)\left[a dt- \left\{ r^2 + (a+n)^2 \right\} d\varphi \right]}{a\rho^2} .\ \label{eq:electromagneticpotential}
\end{eqnarray} 
For $ g,n=0 $, the space-time metric reduces to the case explained in \cite{Xu2018}. One can check that this is the solution of Rastall (modified Einstein) field equation using \textit{Mathematica} package RGTC. The Einstein tensor is shown in \ref{appendixA}. This metric  is more general than the previous metric (\ref{eq:metricresult}) and than the metrics that are shown in several articles \cite{Heydarzade2017,Xu2018,Wang2017a} which also contains magnetic charge and NUT charge (for vanishing cosmological constant). The existence of magnetic charge provides an opportunity to study the axially symmetric black hole solutions with magnetic monopole while the NUT charge is related to the gravo-magnetic monopole \cite{Nouri1997}. More special about NUT charge is that its presence constructs asymptotically non-flat solution, so it will be interesting for further study to compute the mass and angular momentum using Komar integral whether it will be well-defined or not.

We also construct the energy-momentum tensor from the gravitational tensor as 
\begin{equation}
T_{\mu\nu} = \frac{1}{\kappa}\left(G_{\mu\nu} +\kappa\lambda g_{\mu\nu} R \right). \label{eq:emtensor}\
\end{equation}
The non-zero components of this tensor are shown in \ref{appendixB}. Obviously we can see that all non-zero components are dependent of parameters $ a, M, e, g, n, \alpha, \omega$ and $ \kappa\lambda $. As we know, NUT charge is also known as a twisting parameter \cite{Badawi2006}. With vanishing spin, the momentum density component of energy-momentum tensor does not vanish. However, the magnetic charge has similar effect as electric charge in this matter tensor but it dyonically charges the solution. Regarding the discussion of gravo-magnetic monopole \cite{Nouri1997} of NUT parameter, we may expect that our solution contains two magnetic charges. These charges come from the electromagnetic theory and gravitational theory which give another feature of magnetic monopole. It is believed also that in the early universe, the magnetic monopole existed. The distribution of the quintessential matter around black hole solution (\ref{eq:metricresult1}) is also determined by all parameters especially NUT charge and Rastall parameter that give more terms. 

\subsection{Rastall gravity is not equivalent with Einstein gravity}

One of the analysis of Visser \cite{Visser2018} to say that Rastall gravity and Einstein gravity are equivalent is that the term containing Rastall parameter is just a redefinition of cosmological constant term. For vacuum equation he shows that $ \frac{1}{4}\lambda R g_{\alpha\beta} = 0 $ for $ \lambda \neq 1 $ and $ \frac{1}{4}\lambda R g_{\alpha\beta} = -\Lambda g_{\alpha\beta} $ for $ \lambda = 1 $.  When matter is added, he shows that $ \frac{1}{4}\lambda R g_{\alpha\beta} = -\frac{\lambda}{4\lambda -4} T_R g_{\alpha\beta} $ for $ \lambda \neq 1 $ and $ \frac{1}{4}\lambda R g_{\alpha\beta} = -\Lambda g_{\alpha\beta} $ for $ \lambda = 1 $ where $ T_R $ is the trace of Rastall's energy momentum tensor used by Visser and in his paper, he uses normalization $ \kappa =1/4 $. If this argument is right, the solution of those conditions should represent a metric tensor containing cosmological constant or, at least, terms that can be redefined as a cosmological constant. Nonetheless, for example, the following Kerr-NUT-dS(AdS) black hole
\begin{eqnarray}
ds^2 &=& -\frac{\Delta_r}{\Xi^2\rho^2}\bigg[ dt- \{a\sin^2\theta -2n(1-\cos \theta)\} d\phi \bigg]^2 + \frac{\rho^2}{\Delta_r}dr^2 +\frac{\rho^2}{\Delta_\theta}d\theta^2 \nonumber\\
& & \frac{\Delta_\theta \sin^2 \theta}{\Xi^2 \rho^2}\bigg[adt -\{ r^2 +(a+n)^2 \} d\phi \bigg]^2 , \
\end{eqnarray}
where
\begin{eqnarray}
\Delta_r = r^2 - 2Mr+a^2-n^2 -\frac{r^2(r^2+6n^2+a^2)}{l^2} -\frac{3n^2(a^2 - n^2)}{l^2}, \nonumber\\
\Delta_\theta =  1+ \frac{a\cos\theta(4n+a\cos \theta)}{l^2}, ~~~
\rho^2 = r^2 + (n+a\cos\theta)^2, ~~\Xi = 1+\frac{a^2}{l^2}, \nonumber\
\end{eqnarray}
could be used to analyze whether those two gravitational theories are equivalent or not because the most important feature is the metric tensor. The reason we choose to compare with the Kerr-NUT-dS(AdS) metric is because this metric contains three important parameters that we need which are spin $ a $, NUT charge $ n $ and cosmological constant $ 1/l^2 $ (or negative for AdS space).

If Rastall gravity is just a redefinition of matter or cosmological constant, it should be obvious that our metric could be described as Kerr-NUT-dS(AdS) metric for arbitrary value of $ \omega,\kappa\lambda $. From this metric, we can see that the cosmological constant is coupled with spin and NUT charge in the functions $ \Delta_r, \Delta_\theta $. Let us take a look at the $ \Delta_r $ first. We may write $ \Delta $ as
\begin{equation}
 \Delta = r^2 - 2Mr+a^2-n^2 -\left(\sum_{n} \alpha_n \right)r^\upsilon, ~~~\upsilon = \frac{1}{1-\sum_{m} \kappa\lambda_m} ,\ 
\end{equation}
where we set $ e,g,\omega=0 $ in this case to simplify the explanation. The constants $ \alpha_n, \lambda_m $ can be identified as any constants to reproduce the function $ \Delta_r $, for example
\begin{eqnarray}
\alpha_0 r^\upsilon= \frac{3n^2(a^2-n^2)}{l^2},~~~ \alpha_1 r^\upsilon =0,~~~ \alpha_2 r^\upsilon = \frac{r^2(6n^2+a^2)}{l^2}, ~~~ \alpha_3 r^\upsilon =0,~~~ \alpha_4 r^\upsilon =\frac{r^4}{l^2} . 
\end{eqnarray}
However, in our metric the function $ \Delta_\theta $ does not appear. It cannot be produced using the redefinition of $ \alpha, \lambda $. Another argument is that $ R \sim T $ is not a constant but it is dependent on the coordinate. Ricci scalar of our solution can be seen in Eq. (\ref{eq:scalarRicci}). So from this first point of view, Rastall gravity is not equivalent to Einstein gravity. Note that for non-linear form of scalar Ricci, as the Gauss-Bonnet gravity or $ f(R)$ gravity, on the non-conservative matter tensor equation (\ref{eq:nonconservedmatter}) might produce the equivalence 
between these two theories. Nevertheless, the non-linear term is beyond the Ratall theory that can be proposed for further study as
\begin{eqnarray}
\nabla_\mu T^{\mu\nu} = \lambda \nabla^\nu K(R), \label{eq:nonconservedmatter}\
\end{eqnarray}
where $ K(R) $ is the function containing non-linear terms of $ R $ \cite{KaiQianChinese2019}.
Related to the energy-momentum tensor (\ref{eq:emtensor}), the role of the Rastall theory also can be seen. If the statement that Einstein gravity is equivalent to Rastall gravity is right, $ \kappa\lambda $ only describes the rearrangement quintessential matter, however it does not. Moreover, Rastall gravity might correspond to Mach principle, if this theory is not equivalent to Einstein gravity. Recently, in \cite{Fisher2019,Moraes2019} it is argued that Rastall gravity is a special case of $ f(R,T) $ gravity. If we also consider the argument in \cite{Fisher2019,Moraes2019}, so once again, Rastall gravity is not equivalent  to Einstein gravity.

\section{Some black holes structures}
\label{sec:structure}
We have explained that our solution is a new solution in Rastall theory. For further understanding, in this section, we discuss the black holes structures of the black hole solution obtained in the previous section. 

\subsection{Black holes horizons}
\label{subsec:horizon}
\begin{figure}
\resizebox{1.0\textwidth}{!}{
  \includegraphics{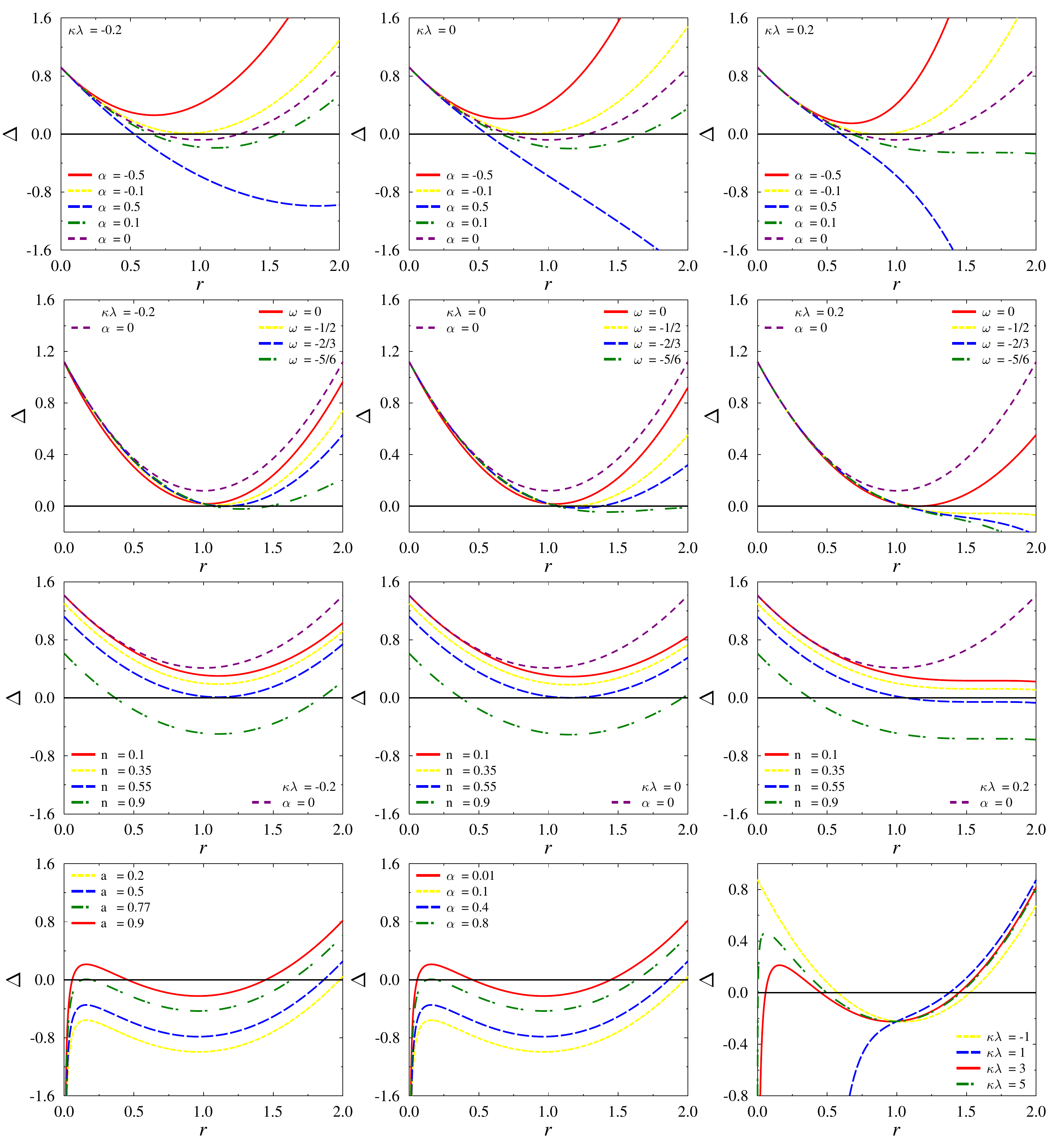}
  }
\caption{The plot showing the behavior of $\Delta $ for $ M=1 $. For the top pictures, we use $ a=1,q=0.64,n=0.7,\omega=-1/2 $ and vary $ \alpha,\kappa\lambda $. The second line pictures, we vary $ \omega,\kappa\lambda $ and use $a=0.8,q=0.8,n=0.4, \alpha=0.1 $, except the purple line $ \alpha=0 $. The third line pictures, $a=1,q=0.65,\alpha=0.1,\omega=-1/2 $ is taken, except the purple line $ \alpha=0 $ and we vary $ n,\kappa\lambda $. The bottom pictures are intended to show a possibility of having three horizons for $ q = 0.7, n = 0.65,\omega=-1/2 $. For left bottom picture, we take $\alpha=0.1,\kappa\lambda=3$ and vary $ a $. For middle bottom picture, we use $a=0.9, \kappa\lambda=3 $ and vary $ \alpha $. The last is right bottom picture for which we take $a=0.9, \alpha=0.1$ and vary $ \kappa\lambda $.} \label{fig:horizon}
\end{figure}

Kerr-Newman-NUT-Kiselev black hole solution in Rastall gravity has the number of the horizon which is dependent on the value of $ \omega $ and $ \kappa\lambda $. Coordinate singularity determines the horizon which corresponds with the space-time. The horizon is a null hypersurface of constant $ r $. Thus, the horizons are the roots of the following equation
\begin{eqnarray}
g_{11}^{-1} = r^2 - 2Mr+a^2+q^2-n^2 -\alpha r^\upsilon =0,  \label{eq:horizon} 
\end{eqnarray}
where $ q^2=e^2+g^2 $. The horizon structure for $ g,n,\kappa\lambda =0 $ is shown in \cite{Wang2017a} but here we provide the more general solution. Performing similar fashion in \cite{Toshmatov2017,Xu2018}, it is found that for general value of $ \omega $, the quintessential intensity satisfies
\begin{eqnarray}
0<\alpha \leq \frac{2-6\kappa \lambda(1+\omega)}{1-3\omega}2^{\frac{3\omega -3\kappa \lambda(1+\omega)}{1-3\kappa \lambda(1+\omega)}},
\end{eqnarray}
where we have put $ M=1 $. For $ \kappa \lambda =0 $, it reduces to the result obtained in \cite{Wang2017a}. Note that even the NUT charge is present, it does not affect the constraint of the quintessential intensity as the spin \cite{Wang2017a}. Because of the existence of quintessence and besides the inner and outer horizons, our black hole solutions might possess a so-called cosmological horizon but it differs from the cosmological horizon when the cosmological constant is non-vanishing. However, it might exist only one horizon or naked singularity. Each horizon radius of the black holes will depend on $a,q,n, \alpha, \omega, \kappa\lambda$ and $M$. In Fig. \ref{fig:horizon}, we show the possibilities of the number of the horizon. 

The top three pictures shows the variation of $ \alpha $ and $ \kappa\lambda $. Negative or positive value of $ \alpha,\kappa\lambda $ is related to the energy condition of the solution, for instance, the explanation for non-rotating solution is given in \cite{Heydarzade2017}. We compare the horizons for negative and positive $ \kappa\lambda $ and also with the solution in Einstein general relativity ($ \kappa\lambda=0 $). For the given value of each parameter, these pictures show four cases of black hole solutions. Firstly, the red solid line describes a black hole solution with naked singularity because no horizon is shown. The yellow line denotes the existence of extremal black hole solution where the inner and outer horizons coincide each other. The purple line clearly tells us that the black holes only have two horizons, yet it is generic Kerr-Newman-NUT black hole because $ \alpha=0 $. The green line in the left and middle pictures has similar meaning as the purple but not for the right one. The green line in the right picture, together with the blue line in every top picture shows that the black hole solution only has a single horizon.

The second and third line pictures are intended for the similar reasons such as the top pictures. Yet we vary the different parameters which are  $ \omega $ and $ n $. We also compare with the Kerr-Newman-NUT black holes (purple line). If we carefully compare the middle pictures with the right pictures for both variation, we can see that the existence of Rastall coupling constant makes the number of the horizon decreases (see the yellow, blue, green lines). Note that $ \omega=0 $ is for the dust domination and $ \omega=-1/2,-2/3,-5/6 $ is for quintessential field domination. The bottom pictures explain that we can have the black holes with three horizons. On the other hand, there might exist the cosmological horizon ($ r_q $) with $ r_-\leq r_+\leq r_q $. If there exist three horizons, so the probabilities of the extremal case are $ r_-=r_+ $, $ r_-=r_q $, $ r_+=r_q $, and $ r_-=r_+=r_q $. It is clear from the left and middle bottom pictures, the green line shows that it fulfills an extremal case $ r_-=r_+ $. Though Rastall coupling constant could make the number of horizon decreases, we may see from the bottom right picture when $ \kappa\lambda =-1 $, the horizons are only two. However, when $ \kappa\lambda =1 $ or $ \kappa\lambda =3 $, the cosmological horizon then appears.

For some values of $ \omega,\kappa \lambda $, the horizon coincides to have only inner and outer parts or has no cosmological horizon. In this case, some thermodynamic properties are easier to be investigated. That is why in \cite{Wang2017b}, they consider only two roots of the horizon.  For more than two roots of the horizon, it will bring to more complicated computation for example to study the entropy product. We show some circumstances that will give rise to two analytic roots in Table \ref{tab:analyticroot}. For cases $ \omega,\kappa\lambda=0,0;-1/3,-1/2;1/3,0 $, the existence of the quintessence increase the horizon radius because of the positive quintessential intensity $ \alpha $ and \textit{vice versa}. For cases $ -1/3,0;0,1/6 $, $ \alpha $ plays a role as the denominator, so $ \alpha $ should not be equal to $ 1 $ or on the other hand, the horizon radius will be infinite.
\begin{table}
\centering
\caption{Analytical inner and outer horizon for some values of $ \omega, \kappa \lambda $.}
\label{tab:analyticroot}
\begin{tabular}{ l  l }
\hline
\hline\noalign{\smallskip}
 $ \omega, \kappa \lambda $ & Horizon $ (r_\pm) $ \\ \hline
 0,0 &$\left(M+\frac{\alpha}{2}\right)\pm \sqrt{\left(M+\frac{\alpha}{2}\right)^2 +n^2 - a^2 -q^2} $\\
 -1/3,0 &$\frac{M}{1-\alpha} \pm \frac{\sqrt{M^2-(a^2+q^2-n^2)(1-\alpha)}}{1-\alpha}$ \\
 0, 1/6 &$\frac{M}{1-\alpha} \pm \frac{\sqrt{M^2-(a^2+q^2-n^2)(1-\alpha)}}{1-\alpha}$\\
 -1/3, -1/2 & $\left(M+\frac{\alpha}{2}\right)\pm \sqrt{\left(M+\frac{\alpha}{2}\right)^2 +n^2 - a^2 -q^2}$\\
  1/3,0 &$ M \pm \sqrt{M^2+n^2 +\alpha-a^2-q^2}$ \\
\hline\noalign{\smallskip}
\hline
\end{tabular}
\end{table}

\subsection{Ergoregion of the black holes}
\label{subsec:ergoregion}

\begin{figure*}[hbtp]
\centering
\resizebox{0.82\textwidth}{!}{
  \includegraphics{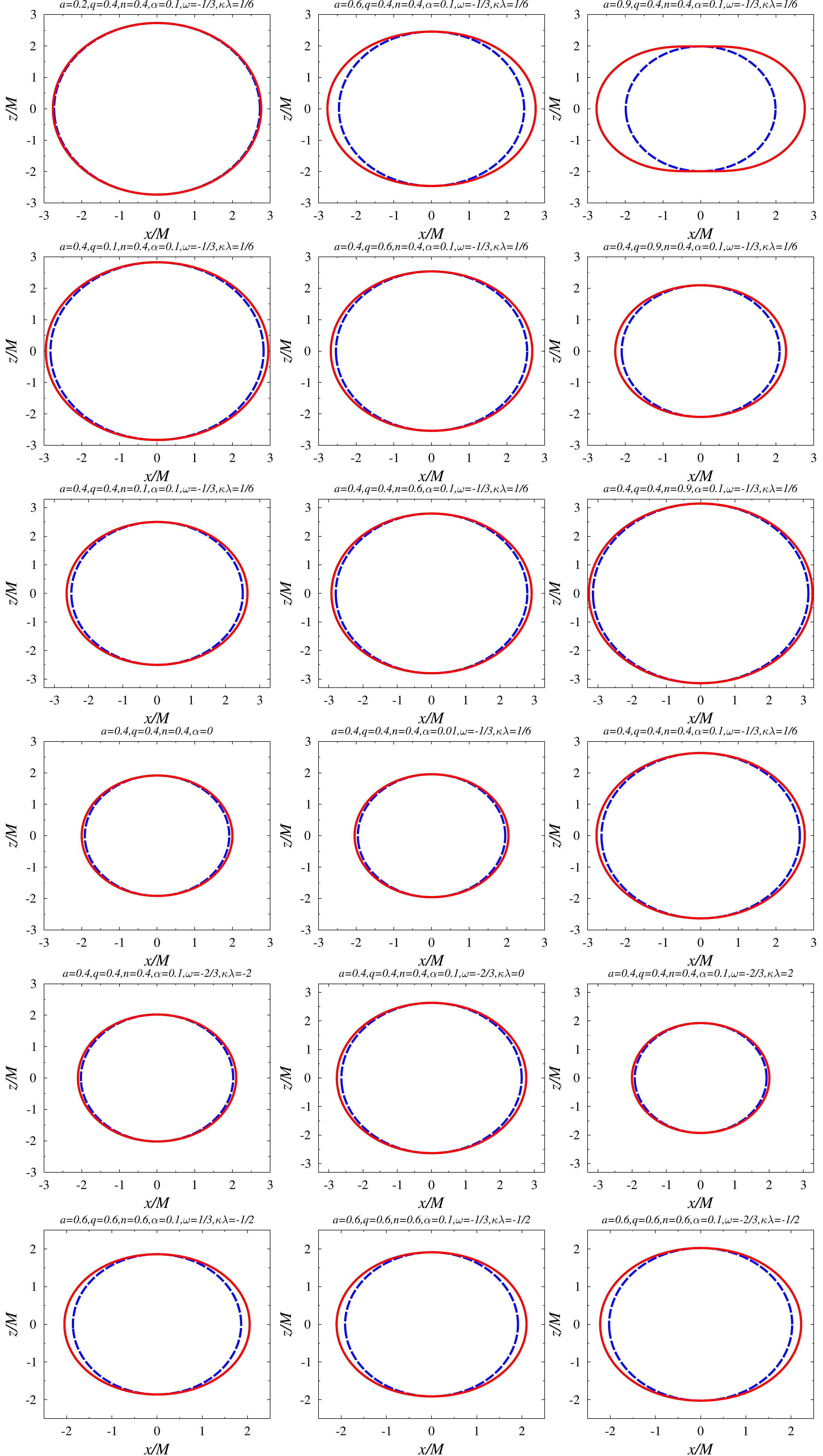}
  }
\caption{The ergoregion of Kerr-Newman-NUT-Kiselev black holes in Rastall gravity with the variation of $ a,q,n,\alpha,\omega,\kappa\lambda.$ The chosen value of each parameter is given on the picture. } \label{fig:ergo1}
\end{figure*}
For rotating black hole, there exists a surface besides the event horizon, i.e. the static limit surface. It corresponds
to the time-like hypersurface where the time translational
Killing vector become null, so it means $ g_{00}=0 $. For the Kerr-Newman-NUT with the quintessence in Rastall gravity, the static limit surface comes from the equation that is given by
\begin{eqnarray}
g_{00}= r^2 - 2Mr+a^2\cos^2\theta +q^2-n^2 -\alpha r^\upsilon =0. \
\end{eqnarray}
These surfaces meet at the poles. The region between them gives rise to ergoregion or ergosphere admitting negative energy orbits where the asymptotic time-like Killing vector becomes space-like. The region of ergosphere is between $ r_+ <r < r_+^{sls} $. It turns out that the shape of the ergosphere also depends on such parameters as the horizon. As an example for the perfect fluid dark matter dominated solution with $ \kappa\lambda =-1/2 $, we have
\begin{eqnarray}
r_+^{sls}=\left(M+\frac{\alpha}{2}\right)\pm \sqrt{\left(M+\frac{\alpha}{2}\right)^2 +n^2 - a^2\cos^2\theta -q^2}.
\end{eqnarray}
The plot of ergosphere is illustrated in Fig. \ref{fig:ergo1} with the variation of all parameters. Note that when the ergoshepere becomes smaller, it means the rotational energy of the black holes is decreasing and \textit{vice versa}.
From Fig. \ref{fig:ergo1}, we can see that the bigger $ a $, the wider ergosphere. But the behavior of the ergosphere due to the presence of $ q $ is different because the bigger charge, the smaller its size. Then if we enlarge the value of NUT charge $ n $ and quintessential intensity $ \alpha $, it makes the ergosphere larger. Remember that $ g $ and $ n $ are related to the magnetic monopole. However, from the ergosphere size, both parameters possesses opposite effect. Then when we vary $ \kappa\lambda $, it is obvious that in Einstein theory the ergosphere is bigger than in Rastall theory because the denominator on the power of $ r $ that contains Rastall coupling constant is equal to $ 1 $. However, for general $ \kappa\lambda \neq 0 $, the increasing of its value causes the increasing of ergosphere's size. Then the bottom pictures show the effect of $ \omega $ for which the decreasing of it makes the ergosphere bigger (clearer when $ \omega=-1/3 $ to $ \omega=-2/3 $).

\section{Some Thermodynamic Properties}
\label{sec:thermo}

In this section, some thermodynamic quantities are investigated. For the black hole solution that is indicated by Eq. (\ref{eq:metricresult1}), the number of horizon depends on $ \omega $ and $ \kappa \lambda $ as the power of $ r $. To study the thermodynamics of this solution, we need to know the event horizon but it is very complicated to find it for non-arbitrary $ \omega, \kappa \lambda $. Hence, here we will define the thermodynamics in term of event horizon $ r_+ $. Firstly, the mass is determined from Eq. (\ref{eq:horizon}) as
 \begin{eqnarray}
M = \frac{1}{2r_+}\left(r_+ ^2 - \alpha r_+^\upsilon +a^2+q^2-n^2 \right). \label{eq:massdef}
\end{eqnarray}

To calculate the Hawking temperature, the tunneling method can be considered as in \cite{Parikh2000,Banerjee2008,Ma2008}. In this fashion, the metric is allowed to be diagonal so, we put $ d\theta=d\varphi=0 $ and take the angular coordinate to be $ \theta=0 $. Hence, the metric and Hawking temperature is expressed respectively as
\begin{eqnarray}
ds^2 = -f(r)dt^2 + \frac{1}{f(r)}dr^2, ~~~~~
T_H = \frac{\partial_r f(r)}{4\pi}\bigg|_{r=r_+}. \ 
\end{eqnarray}
To make use of the tunneling method, we apply the above assumption such that the metric (\ref{eq:metricresult1}) becomes
\begin{eqnarray}
ds^2 = -\frac{\Delta}{r^2 + (a+n)^2}dt^2+ \frac{r^2 + (a+n)^2}{\Delta}dr^2.  \label{eq:metthermo}
\end{eqnarray}
Then for metric (\ref{eq:metthermo}), the Hawking temperature is computed and straightforwardly yields
\begin{eqnarray}
T_H = \frac{2(r_+ - M) - \alpha \upsilon r_+^{\upsilon -1}}{4\pi [r_+^2 + (a+n)^2]}, \label{eq:Hawkingtemp}\ 
\end{eqnarray}
where the mass is given in Eq. (\ref{eq:massdef}). The angular momentum of solution (\ref{eq:metricresult1}) is then
\begin{eqnarray}
\Omega_H = -\frac{g_{03}}{g_{33}}\bigg|_{r=r_+} = \frac{a}{[r_+^2 + (a+n)^2]}. \label{eq:angmomen}\
 \end{eqnarray}
The Coulomb electromagnetic potential is given by \cite{Sakti2018}
\begin{eqnarray}
\Phi_H &=& - K^\mu A_\mu \big|_{r=r_+} = \Phi_e +\Phi_g =  \frac{er_++g(a+n)}{r_+^2+(a+n)^2} -\frac{g(a+n)}{r_+^2+(a+n)^2} ,  \label{eq:coulombpotential} \
\end{eqnarray}
where $ K = \partial _{t}+\Omega_H \partial_{\varphi} $. The resulting Hawking temperature is general for every value of $ \omega, \kappa \lambda $. So for specific matter domination as in \cite{Heydarzade2017}, we just have to put the value of the equation of state. For $ \alpha =0 $, the temperature reduces to Kerr-Newman-NUT black hole solution \cite{Sakti2018}. The form of angular momentum and Coulomb potential is also similar to Kerr-Newman-NUT one but it differs on the definition of $ r_+ $. Then it is obvious that for $ \alpha,n=0 $, it reduces to Kerr-Newman black hole properties. It is worth pointing out that the existence of the quintessence causes another thermodynamic variable to appear \cite{Chen2008,Sekiwa2006}. Hence, we can derive the new thermodynamic variable as
\begin{eqnarray}
\Theta_H =\frac{\partial M}{\partial \alpha}\bigg|_{r=r_+}=-\frac{1}{2}r_+^{\upsilon -1},
\end{eqnarray}
when $ S,J,Q_e,Q_g $ are held constant. We need to add this generalized force in order to satisfy the first law of black hole's thermodynamics \cite{Chen2008,Sekiwa2006}.
 
The area of the black hole is given by
\begin{eqnarray}
A_{BH} &=& \int \sqrt{g_{22} g_{33}} d\theta d\varphi = 4\pi [r_+^2 + (a+n)^2]. \label{eq:areabh} \
\end{eqnarray}
Using the area law of the black hole, the Bekenstein-Hawking entropy can be found as
\begin{eqnarray}
S_{BH} = \frac{A_{BH}}{4} = \pi [r_+^2 + (a+n)^2]. \label{eq:bhentropy}
\end{eqnarray}
For constant value of $ J,Q_e,Q_g $, we can also obtain the heat capacity using the relation that is given by \cite{Maharaj2017,Lee2018,Cai2004}
\begin{eqnarray}
C = \frac{dM}{dT_H}\bigg|_{r=r_+} = \frac{dM/dr_+}{dT_H/dr_+}\bigg|_{r=r_+}, \label{eq:heatcap}\
\end{eqnarray}
Hence from Eq. (\ref{eq:heatcap}), we find that
\begin{eqnarray}
C = \frac{2\pi [r_+^2 + (a+n)^2]^2 }{[r_+^2 + (a+n)^2] \delta -2r_+^2}, ~~\text{where}~~\delta = \frac{r_+^2 - \alpha(\upsilon -1)^2 r_+^\upsilon + a^2 +q^2 -n^2}{r_+^2 - \alpha(\upsilon -1) r_+^\upsilon - a^2 -q^2 +n^2} . \label{eq:heatcap1}\
\end{eqnarray}

The dependence of $ \alpha, \omega $ and $ \kappa \lambda $ on area and entropy is implicitly contained on $ r_+ $ while on heat capacity, it is obviously seen in Eq. (\ref{eq:heatcap1}). The thermodynamic stability of any system is related to the sign of the heat capacity \cite{SaktiPrihadi2019}. For black holes, when the heat capacity is positive $ (C>0) $, it will be stable. Whereas when the heat capacity is negative $ (C<0) $, the black holes are said to be unstable. On the Kerr-Newman-NUT-Kiselev black hole solution in Rastall gravity, the heat capacity is dependent on many parameters. So, to find the thermodynamic stability, those parameters need to be determined. Note that numerator of the heat capacity is always positive so the stability is determined by the sign of the denominator.  It can be obviously seen that the Hawking temperature, entropy, angular momentum, Coulomb potential, and heat capacity are explicitly dependent of NUT charge.

We give examples of the entropy for some arbitrary values of $ \omega, \kappa\lambda $. The values are chosen from the Table {\ref{tab:analyticroot} where the event horizons are provided. Then the extremal Bekenstein-Hawking entropies are given by
\begin{eqnarray}
S_{BH}^{(1)} &=& \pi \left[\left(M+\frac{\alpha}{2}\right)^2 +(a+n)^2 \right], \label{eq:entropy3}\\
S_{BH}^{(2)} &=& \pi \left[\left(\frac{M}{1-\alpha}\right)^2 +(a+n)^2 \right], \label{eq:entropy2}\
\end{eqnarray}
for the black hole solution with perfect fluid dark matter domination $ \omega,\kappa\lambda =-1/3,-1/2 $ and dust domination $ \omega,\kappa\lambda =0,1/6 $, respectively. The investigation of every specific value of $\omega,\kappa\lambda $ is studied in \cite{Heydarzade2017} for non-rotating case related to the violation of the strong energy condition. We can note that in Eq. (\ref{eq:entropy2}), the quintessential intensity is restricted on $ \alpha = 1 $ to gain finite entropy for the given value of $ \omega, \kappa \lambda $. We will also have the same value of entropy for $ \omega,\kappa\lambda =0,0 $ and $ \omega,\kappa\lambda =-1/3,0 $ for dust domination and perfect fluid dark matter domination in Einstein theory, respectively. When the quintessence is absent, it reduces to the extremal Kerr-Newman-NUT black hole's entropy.

\section{Central charge from CFT}
\label{sec:central}

The Kerr/CFT correspondence, firstly proposed in \cite{Guica2009}, has been successful to compute the microscopic origin of extremal black hole entropy. To find the microscopic entropy, the famous Cardy formula is applied that is given by
\begin{eqnarray}
S_{CFT} = \frac{\pi ^2}{3}(c_L T_L+c_R T_R),\label{eq:cardyentropi}\
\end{eqnarray}
where $ c_L,c_R $ are the left-moving and right-moving central charges while $ T_L,T_R $ are the left-moving and right-moving conformal temperatures. Using this correspondence, it is proved that entropy of extremal black holes from CFT (\ref{eq:cardyentropi}) agrees with the Bekenstein-Hawking one \cite{Hartman2009}-\cite{Sakti2018}. In the previous section, the Bekenstein-Hawking entropy for Kerr-Newman-NUT-Kiselev black hole in Rastall gravity has been calculated. However, the microscopic origin of this entropy (\ref{eq:bhentropy}) is not calculated yet using the Kerr/CFT correspondence. One reason is because the canonically
conserved charge for Rastall theory of gravity is not calculated yet. But, for the other modified gravity theory can be found in \cite{Adami2017}. The asymptotic charge that is related to the asymptotic symmetry is used to find the central charge. The central charge shows up from the central extension that comes from the Dirac bracket of asymptotic charges \cite{Barnich2002,Barnich2008}. In this section, we are going to find the central charge considering the Kerr/CFT correspondence that relates entropy from CFT and the Bekenstein-Hawking entropy. We argue that this correspondence may be implemented because when $ \kappa\lambda $ vanishes, it becomes the usual Einstein-Maxwell system with the scalar field. In \cite{Skanata2012,Ghezelbash2013,Ghezelbash2014,SaktiGhezelbash2019}, this relation is also performed for deformed black hole solutions because the central charge is not calculated yet. For the solutions consisting of the electromagnetic and scalar field besides the gravity, this correspondence is able to prove the relation between the Bekenstein-Hawking entropy and CFTs entropy such as for Kerr-Sen black hole solution  \cite{Ghezelbash2009} and black hole with dilaton-axion field \cite{Li2010}. Generally, this relation is valid not only for extremal black holes, but also for non-extremal black holes \cite{Castro2010}-\cite{SiahaanAcc2018}.

Before we derive the central charge, it is common to show the isometry of the near-horizon extremal black hole solutions. In addition, the isometry should be $ U(1)\times SL(2,R) $ to employ the asymptotic symmetry group. So we will show that the near-horizon extremal Kerr-Newman-NUT-Kiselev black hole solutions in Rastall gravity have that isometry. Firstly, we need to define the following coordinates transformation based on \cite{Guica2009,Compere2017} which is given by
\begin{eqnarray}
r = r_+ + \epsilon r_0 y, ~ t = \frac{r_0}{\epsilon}\tau, \varphi = \phi +\Omega_H \frac{r_0}{\epsilon}\tau , \label{eq:coortransmatterdom}\
\end{eqnarray}
where $ r_0^2 = r_+^2 +(a+n)^2 $. After taking $ \epsilon \rightarrow 0 $, the near-horizon extremal metric of (\ref{eq:metricresult1}) reads
\begin{eqnarray}
ds^2 &=&\frac{\rho_+^2}{V} \left(-y^2d\tau^2 + \frac{dy^2}{y^2}+ V d\theta ^2 \right)+ \frac{r_0^4 \sin^2\theta}{\rho_+^2} \left( d\phi + \frac{2ar_+}{V r_0^2} yd\tau \right)^2 ,\label{eq:RDnearhor3}\
\end{eqnarray}
where  $ \rho_+^2 = r_+^2 +(n+a \cos\theta)^2 $. Straightforwardly, we could perceive that the near-horizon metric (\ref{eq:RDnearhor3}) has $ AdS_2\times S^2 $ topology. It possesses an isometry $ U(1)\times SL(2,R) $ which is generated by the following vector fields
\begin{eqnarray}
\zeta_0=\partial_{\phi}, ~~~~~ X_1 = \partial_\tau, ~X_2 = \tau \partial_\tau - y \partial_y , ~ X_3 = \left(\frac{1}{2y^2}+\frac{\tau^2}{2} \right)\partial_\tau -\tau y \partial_y - \frac{2ar_+}{Vr_0^2y}\partial_\phi .\label{eq:isometrynearhorizon}
\end{eqnarray}

That isometry exhibits that Kerr/CFT should be valid for our extremal solutions. Furthermore, to obtain the central charge, we need to compute the the conformal temperatures. In computing this temperatures, we apply the fashion shown in \cite{Sinamuli2016,Compere2017,Sakti2018} for which it is given by
\begin{eqnarray}
 T_R = \frac{T_H r_0}{\epsilon}\bigg|_{r=r_+} , ~~~~~ T_L = - \frac{\partial T_H/\partial r_+}{\partial \Omega_H / \partial r_+}\bigg|_{r=r_+} ,\ \label{eq:generalCFTtemperature}
\end{eqnarray}
where $ r_0 $ is a constant to factor out the overall scale of the near-horizon geometry and $ \epsilon  $ is an infinitesimal constant for extremal black holes. It is clear that for extremal black holes, the right-moving temperature will vanish because it is proportional to the Hawking temperature. Hence, the remaining temperature that is required to be calculated is the left-moving temperature based on Eq. (\ref{eq:cardyentropi}). Using the formula of left-moving temperature in Eq. (\ref{eq:generalCFTtemperature}), we finally derive 
\begin{eqnarray}
T_L = \frac{[2-\alpha \upsilon(\upsilon -1)r_+^{\upsilon -2}][r_+^2 + (a+n)^2]}{8\pi ar_+}. \
\end{eqnarray}
The conformal temperatures come from the construction of the Frolov-Thorne vacuum for generic rotating black holes. When $ \alpha=0 $, it reduces to the left-moving temperature of Kerr-Newman-NUT black hole. After getting the left-moving temperature, the Cardy formula (\ref{eq:cardyentropi}) is applied. Since we consider $ S_{CFT}=S_{BH} $, finally the left-moving central charge is given as follows
\begin{eqnarray}
c_L = \frac{24ar_+}{2-\alpha \upsilon(\upsilon -1)r_+^{\upsilon -2}} .\label{eq:centralcharge}
\end{eqnarray}
In general, the central charge and left-moving temperature show up as a function of the quintessential intensity, equation of state, and Rastall parameter besides the other parameters. It is interesting because any specific of matter domination as the quintessential model, along with the Rastall coupling constant affects the value of these quantities. It is obvious that $ \alpha \upsilon(\upsilon -1)r_+^{\upsilon -2} $ may not be equal to $ 2 $ to produce finite left-moving central charge. So the value of $ \omega,\kappa\lambda $ are required to be constrained but we will not explain in more detail within this paper.

As the entropy, we give examples of the central charge for some arbitrary values of $ \omega, \kappa\lambda $ given in the Table {\ref{tab:analyticroot}. We find the following central charges
\begin{eqnarray}
c_L^{(1)} &=& 12a\left(M+ \frac{\alpha}{2} \right), \\
c_L^{(2)} &=& \frac{12aM}{(1-\alpha)^2},\
\end{eqnarray}
for the black hole solution with perfect fluid dark matter domination $ \omega,\kappa\lambda =-1/3,-1/2 $ and dust domination $ \omega,\kappa\lambda =0,1/6 $, respectively. We will also have the same value of central charge for $ \omega,\kappa\lambda =0,0 $ and $ \omega,\kappa\lambda =-1/3,0 $, respectively. When the quintessential intensity vanishes, the central charge coincides to the central charge of extremal Kerr-Newman-NUT black hole \cite{Sakti2018}. In addition, when $ q $ vanishes, it reduces to the result obtained in \cite{Ghezelbash2012,Sakti2018a}.

\section{Summary}
\label{sec:summary}

By performing the Demia{\'n}ski-Newman-Janis algorithm, we obtained Kerr-Newman-NUT-Kiselev black hole solutions in Rastall theory of gravity which is the extension of Kerr-Newman-Kiselev black holes in Rastall theory. This new solution contains magnetic charge and NUT parameter which are related to the magnetic monopole. It is believed that in the early universe, the magnetic monopole existed. The horizon analysis and ergosphere were also studied to see the possibilities of having several horizons and the dependence of each parameter. For some arbitrary values of $ \omega, \kappa\lambda $, this solution coincides to possess inner and outer horizons. The size of the ergosphere is dependent of each parameter related to the rotational energy of the black hole. We also could map several values of $ \omega $ in Rastall gravity to the other values of  $ \omega $ in Einstein general relativity. But we did not say that these two theories are equivalent because Rastall theory seems more Machian and needs to face the challenges of the observational evidence. From the argument that we pointed out, we believed that Rastall gravity is not equivalent to Einstein gravity.

Several thermodynamic properties were investigated. We defined the mass from the horizon equation. It is useful to define the Hawking temperature that we computed by applying tunneling method. The angular momentum, Coulomb electromagnetic potential and heat capacity were also calculated. Because of the quintessential field, we needed to add the generalized force in order to satisfy the first law of black hole's thermodynamics. Then the Bekenstein-Hawking entropy was also derived which is equal with the area law of the black hole. It could be seen that the dependence of the quintessential intensity $ \alpha $ implicitly lies on the horizon $ r_+ $ but for the temperature, its dependence is obvious. The Hawking temperature, entropy, angular momentum, Coulomb potential, and heat capacity are explicitly dependent of NUT charge also, besides the spin.
It will be interesting to study the thermodynamic stability of this solution from the heat capacity, but we did not explain in this article.

Instead of using the asymptotic symmetry group in Kerr/CFT correspondence, we calculated the central charge of the extremal black hole solutions by equating the entropy from the CFT with the Bekenstein-Hawking one which was already proved. Before finding the central charge, we showed the isometry of the near-horizon extremal form of our solution. Furthermore, the conformal temperatures were computed first where the temperatures that we needed are right- and left-moving temperatures. But from the fact that the right-moving part was proportional to the Hawking temperature, it remained zero in extremal case. Finally, the remaining central charge was obtained. The left-moving central charge depends on the quintessential intensity explicitly. Generally, $ \alpha \upsilon(\upsilon -1)r_+^{\upsilon -2} $ may not be equal to $ 2 $ to produce finite left-moving central charge. The central charges for $ \omega,\kappa\lambda =-1/3,-1/2 $ and $ \omega,\kappa\lambda =0,1/6 $ were also shown as the examples. In the future, we want to study this solution using the famous Kerr/CFT correspondence to confirm the result that we obtained in this article.

\section*{Acknowledgments}

We gratefully acknowledge the support from Ministry of Research, Technology, and Higher Education of the Republic of Indonesia and PMDSU Research Grant. We also want to express many thanks all members of Theoretical Physics Laboratory, Institut Teknologi Bandung for the valuable support.


\appendix
\section{Einstein tensor of Kerr-Newman-NUT black holes with quintessence in Rastall gravity}
\label{appendixA}
We derive the Einstein tensor of Kerr-Newman-NUT black holes with the quintessence in Rastall gravity using \textit{Mathematica} package RGTC as
\begin{eqnarray}
G_{tt} &=& \frac{2\left(r^4 -2r^3 +a^2r^2 -a^4 x^2(1-x^2) \right)W'}{\rho^2}-\frac{ra^2\sin ^2 \theta W''}{\rho^4} \nonumber\\
& -& \frac{2n \left[n\left\{r^2 +a^2(1-x^2)\right\} + 2a^3 x(1-x^2) \right]}{\rho^6}, \
\end{eqnarray}
\begin{eqnarray}
G_{t\phi} &=& \frac{4nr^2(1-x)(a^2 -\Delta)W'}{\rho ^2} +  \frac{a(1-x^2)(r^2+n^2)^2 X}{2r^2\rho ^6}  \nonumber\\
&+&\frac{2ar^2(1-x^2)(2r(M-r)-(q^2 -\alpha r^\upsilon))W'}{\rho ^6} \nonumber\\
&+ & \frac{an(1-x)\left[4r^4(x-2)W' + (1-x)(n^2+r^2) X \right]}{r^2\rho ^6} \nonumber\\
& +& \frac{a^3(1-x^2)\left[(1+x^2) X-8r^2W' \right]}{2\rho ^6} + \frac{a^3 n^2 (1-x^2)(1+4x+x^2)X}{2r^2 \rho ^6}\nonumber\\
\
\end{eqnarray}
\begin{equation}
G_{rr} = -\frac{2r^2 W'}{\rho^2 \Delta}, 
\end{equation}
\begin{eqnarray}
G_{\theta\theta} &=& - \frac{2a^2 x^2 W'}{(1-x^2)\rho^2} -\frac{2(n^2+2anx)W'}{(1-x^2)\rho ^2} - \frac{rW''}{(1-x^2)},\
\end{eqnarray}
\begin{eqnarray}
G_{\phi\phi} &=& - \frac{a(1-x^2)(r^2+a^2)(a^2+(2r^2+a^2)(2x^2-1) W'}{\rho^2} \nonumber\\
& -& \frac{2r^3a(1-x^2)^2 W W'}{\rho^2} + \frac{r^2a^2(1-x^2)^2(r^2+a^2-n^2-\Delta)W}{\rho ^6} \nonumber\\
& + &\frac{a^2(r^2+a^2)(1-x^2)\left[2a^2x^2+r^2(4x^2-2)\right] W'}{\rho ^6} \nonumber\\
&+& \frac{\Delta (x-1)^2 (2n+a+ax)^2(2r^2W'+\rho^2 X)Y}{2r^2\rho ^8} + \frac{a^3 x(2n+2nx+ax)X}{2r^2\rho ^8} \nonumber\\
&-& \frac{\Delta(x-1)^2(2n+a+ax)^2(r^2+n^2)(2r^5W''+n^2 X)}{2r^2\rho ^8} \nonumber\\
&+& \frac{a[4r^2 n(x-2)W' +(1+x)X]}{\rho^8} +\frac{a^2\left[r^2(1+x^2)+n^2(1+4x+x^2) \right]}{2r^2\rho ^8} \nonumber\\
&+& \frac{(r^2+n^2)(1-x^2)(2r^5 W'' +n^2x)Y^2}{2r^2 \rho ^8} + \frac{2r^2a(x^2 -2)W'}{\rho^8} .\
\end{eqnarray}
where $ x = \cos \theta $, $ W = M - \frac{q^2}{2r^2}+ \frac{\alpha}{2}r^{\upsilon -1} $, $ q^2 =e^2+g^2 $, $ X = \alpha \upsilon (\upsilon -1)r^\upsilon $ and $ Y = r^2 + (a+n)^2 $. The Ricci scalar is
\begin{equation}
R =\frac{X}{r^2 \rho^2}. \label{eq:scalarRicci}
\end{equation}
Note that the total gravitational tensor is 
\begin{equation}
G_{\mu\nu}+\kappa\lambda g_{\mu\nu}R . \
\end{equation}

\section{Energy-momentum components of Kerr-Newman-NUT black holes with quintessence in Rastall gravity}
\label{appendixB}

Using the gravitational tensor, we may construct the energy-momentum tensor of the Kerr-Newman-NUT black holes solution with the quintessence in Rastall gravity. We can get the following equation
\begin{equation}
T^\alpha_\nu = \frac{1}{\kappa}\left(G^\alpha_\nu +\kappa\lambda \delta^\alpha_\nu R \right),
\end{equation}
by rising one index from Eq. (\ref{eq:emtensor}). Finally, we can find
\begin{equation}
E =T^t_t = - \frac{2\left(r^4 + r^2(2a^2 +n^2)+4an -r^2 a(a+2n)x \right)}{\kappa\rho ^6} -\frac{a(x-1)(a+2n+ax)X}{2r^2\kappa\rho^4} + \frac{\lambda X}{r^2 \rho^2},\
\end{equation}
\begin{equation}
P_r = T^r_r = -\frac{2r^2 W'}{\kappa\rho^4} + \frac{\lambda X}{r^2 \rho^2},
\end{equation}
\begin{equation}
P_\theta = T^\theta_\theta = \frac{2r^2 W'}{\kappa\rho^4}-\frac{X}{2r^2\kappa\rho^2} + \frac{\lambda X}{r^2 \rho^2},
\end{equation}
\begin{eqnarray}
P_\phi = T^\phi_\phi &=&  -\frac{r^3(r^2+n^2)W''}{\kappa\rho^6}-\frac{2ar^2[a(x^2-2)+2n(x-2)]W'}{\kappa\rho^6} -\frac{a^3x(2n+ax+2nx)X}{2r^2\kappa\rho^6}  \nonumber\\
& - &\frac{(r^2+n^2)[n^2+2an(1+x)]X}{2r^2\kappa\rho^6}-\frac{a^2[r^2(1+x^2)+n^2(1+4x+x^2)]}{2r^2\kappa\rho^6}+ \frac{\lambda X}{r^2 \rho^2} ,\
\end{eqnarray}
\begin{equation}
S = T^\phi_t = -\frac{4ar^2 W'}{\kappa\rho^6}+ \frac{aX}{2r^2\kappa\rho^4}.
\end{equation}



 \bibliographystyle{elsarticle-harv}

\end{document}